\documentclass[aps,floats,amssymb,twocolumn,prb,figure]{revtex4-1}
\usepackage{calc}
\usepackage{psfrag}
\usepackage{graphicx}
\usepackage{color}
\usepackage{pgfplots}
\usepackage{float}
\usepackage{caption}
\usepackage[export]{adjustbox}
\usepackage{wrapfig}
\usepackage{pgfplots,wrapfig}

\usepackage{tikz}
\usepackage{subcaption}

\usepackage[compatibility=false]{caption}
\usepackage[skip=20pt,font=scriptsize]{caption}
\usepackage{lipsum}
\usepackage{mwe}

\begin{document}
\captionsetup[subfigure]{position=top, textfont=normalfont,singlelinecheck=off,justification=raggedright}

\title{Excitonic physics in a Dirac quantum dot}

\author{ V. Raca$^{1}$ and M. V. Milovanovi\'c$^{2}$}
\affiliation{$^{1}$ Faculty of Economics, University of Belgrade, 11001 Belgrade, Serbia}
\affiliation{$^2$ Scientific Computing Laboratory, Center for the Study of Complex Systems, Institute of Physics Belgrade, University of Belgrade, Pregrevica 118, 11080 Belgrade, Serbia}

\begin{abstract}
We present a description of vacuum polarization in a circular Dirac quantum dot in two spatial dimensions assuming
$\alpha$ - the relative strength of the Coulomb interaction small enough to render an approximation with a single
electron (hole) lowest energy level relevant. Applying this approximation, we find that for $ \alpha_c \approx 1.05$ the
lowest level is half-filled irrespective of the number of flavors that are present. The ground state can be represented as a superposition of particular (even number) excitonic states which constitute an excitonic cloud that evolves in a crossover manner. The ground state is degenerate with an intervalley excitonic state at $ \alpha_c
\approx 1.05$, a critical strength, that in our approximation marks a point with single electron and exciton resonances.
\end{abstract}

\maketitle

\section{Introduction}

The experimental discovery of graphene [\onlinecite{nov1}], a two-dimensional hexagonal lattice of carbon atoms, stirred the interest for Dirac physics in condensed matter theory. According to band-structure calculations, in graphene there are two special points where the valence and conduction band  meet, and make two Dirac cones with linear dispersion. This is the basis for the explanation of a peculiar (particle-hole symmetric) integer quantum Hall effect in graphene [\onlinecite{nov2}] - a prominent feature of the quasirelativistic behavior.  But the effect will be still in place if we assume a mass term in the Dirac description [\onlinecite{sg}]. If we neglect any presence of an explicit symmetry breaking that will induce a mass term, we still need to consider the possibility for spontaneous symmetry breaking  due to electron-electron (Coulomb) interactions. Then the underlying mechanism for the presence of a band gap (a mass term) would be a creation  of (due to the Coulomb interaction) bound particle-hole pairs i.e. excitons. In graphene the importance of the Coulomb interaction is measured by the effective ``fine structure constant" i.e. $\alpha = e^2/(\kappa \hbar v_F)$, where $e$ is the unit of electric charge, $\hbar = h /(2 \pi)$, with $h$, Planck's constant, $v_F$ is the Fermi velocity in graphene, an analog of the speed of light $c$, $v_F \approx c/300 $, and $\kappa$ represents the dielectric constant of the surrounding medium. We have $ \alpha \approx 2.16 /\kappa $, where $\kappa = 1$ in vacuum (i.e. $\alpha_{vac} \approx 2.16$) or, if a graphene sheet is sandwiched between two different media, it equals the average of the corresponding dielectric constants. Typically, graphene transport experiments are performed on a $SiO_2$ substrate with $\alpha_{SiO_2} \approx 0.79$. Thus $\alpha$, the dimensionless measure of the effectiveness of the Coulomb interaction, is much larger, $\alpha = e^2/(\kappa \hbar v_F) \sim 1$, in graphene, than in the (real) relativistic (quantum electrodynamics) case, for which $\alpha = e^2/(\kappa \hbar c) = 1/137.$ Therefore, electrons and holes are strongly attracted to each other in graphene near Dirac points. Theoretical investigations [\onlinecite{kh,dr,gu,wa}] lead to an expectation that a mass gap would exist in a suspended graphene, i.e. for a sufficiently large $\alpha$, $\alpha \sim 2$. In the experiment of Ref. [\onlinecite{mo}] on suspended graphene, no sign of an insulating state is observed down to $1 K$, and, in the reference, the authors state a conservative estimate on any possible bandgap as $ < 0.5  meV$.

The large value of $\alpha$ in graphene leads to many phenomena connected and related to the vacuum polarization i.e. an instability towards creation of particle-hole pairs - excitons of interacting Dirac fermions. Our focus  here would be the exciton physics (vacuum polarization) in a circularly confined Dirac system. In this context we would like to mention  well explored phenomena connected  with the presence of charged impurities in graphene [\onlinecite{sh1,sh2}]. The charge  of the impurity, $Z e$, need not to be large $(Z \gtrsim 1)$ to lead to ``supercritical atomic collapse", which in the case of (non-interacting) graphene leads to an infinite family of quasi-bound states
[\onlinecite{sh1}] (although massless particles cannot form bound states). What is of a particular interest of us here, is the effect of vacuum polarization in which, through exciton formation the impurity is partially screened with a long polarization (screening charge) tail. The algebraic decay of the polarization reflects the absence of any length scale in the Dirac equation [\onlinecite{gu}]. In Refs. [\onlinecite{ga,wa}] it was emphasized that a close connection exists
between the supercritical Coulomb center problem [\onlinecite{sh1,sh2}] and the many-body excitonic instability in the interacting graphene case, which are both followed by the vacuum polarization phenomenon [\onlinecite{ko,kor}].


In this paper we will discuss a finite size Dirac system  i.e. a quantum dot, taking into account the simple fact that, due to the relativistic spectrum, the lowest energy level, which energy we will denote by $\epsilon_1$, is strongly affected by the Coulomb interaction, $V$, i.e.  $V \sim \epsilon_1$. Namely, both $V$ and $\epsilon_1$, scale with the only length present, the radius of the dot, $R$, as $V \sim \epsilon_1 \sim 1/R$. We assumed $V \sim \epsilon_1$, because the dimensionless constant $\alpha$ that measures the relative strength of the Coulomb interaction (as previously discussed) is of the order one in graphene, $\alpha \sim 1$.

As expected, interactions in these small Dirac systems can not cause dramatic effects as shown  in Ref. \onlinecite{pe}. But, especially close or at the neutrality point of the Dirac system, as we will show, we can track down the excitonic physics - the creation of particle-hole pairs from the vacuum in a very efficient manner. Assuming that only the lowest energy state is affected by the Coulomb interaction, but taking all excitonic physics exactly, we were able to reach conclusions for arbitrary number of fermionic flavors, $N_f$. In the case of the graphene, with the spin degree of freedom, $N_f = 2$.  We find that, irrespective of $N_f$, there is a critical $\alpha$, $\alpha_c \approx 1.05$, for which the lowest energy level is half-filled with excitons. That is the critical strength of the Coulomb interaction, after which the polarization (the number of
excitons) increases monotonously, but the ground state does not experience a crossing, and evolves in a crossover manner from vacuum (no excitons). At $\alpha_c$ degeneracies of the levels of the reduced subspace occur, that include the ground state which becomes degenerate with  multiplets of states with odd number parity of excitons. Thus $\alpha_c$ can signify a characteristic value of the Coulomb interaction for an exciton resonant transport. Also we will present the energy necessary to bring electron to the neutral system (``charging energy") as a function of $\alpha$, which can be relevant in the spectroscopic measurements of graphene quantum dots - see the latest important work in that direction
in Refs. \onlinecite{leeetal,guietal,qiaoetal}.

In the literature we can find various proposals for graphene quantum dots. Constructions with different one-dimensional potentials, which overcome Klein paradox by considering states with finite momentum in the transverse direction (strip geometry) were proposed and analyzed in Refs. \onlinecite{s1,s2,s3}. Similarly, the circular geometry leads to quasi-bound states [\onlinecite{b1,b2,b3,b4,mp,b6}], and may even lead to bound states if a confining potential enters as a mass term in the Dirac equation  [\onlinecite{bm}].  Here we will use the circular  set-up with a confining mass term that was used in Ref. \onlinecite{pe}, which analyzed the situation slightly away from the neutrality point (charged systems), with necessary approximations in that case.

The rest of the paper is organized as follows: in the first part of Section II we review details of the set-up and solutions for the non-interacting circular Dirac dot [\onlinecite{pe}], and, then, in the second part, we describe the approximation that we apply (in order to deal with the Coulomb electron-electron interaction), and its consequences. The complete solutions for the cases with the number of flavors equal to $N_f = 1$, and $N_f = 2$, are given in the Section II and Appendices A and B. Section III is devoted to a discussion and conclusions.

\section{Dirac quantum dot}

In the following we will introduce the model of the Dirac quantum dot that we consider. We begin with $2 \times 2$ Dirac Hamiltonian with a mass term,
\begin{equation}
H = v_F \vec{\sigma} \vec{p} + \mu (\vec{r}) \sigma_z ,
\end{equation}
where $\vec{\sigma} = (\sigma_x , \sigma_y )$ and $\sigma_z$ are Pauli matrices, by $\mu$ we denoted a mass, and $v_F$ is the Fermi velocity as before. As shown in, for example, Ref. \onlinecite{bm}, the unitary operator $ U = \sigma_x $ together with $K$, the complex conjugation, transforms the Hamiltonian, $H$, into
\begin{equation}
H' = U K H K U^{-1} = - H,
\end{equation}
and thus if
\begin{equation}
\Psi  = \left[
  \begin{array}{c}
   \Psi_1 \\
   \Psi_2 \\
  \end{array}
\right] \label{eigenstate}
\end{equation}
is an eigenstate of $H$ with energy $E$, then
\begin{equation}
\Psi'  = \left[
  \begin{array}{c}
   \Psi_2^* \\
   \Psi_1^*\\
  \end{array}
\right],
\end{equation}
is also an eigenstate with energy $ - E$. If we consider time-reversal, $U$ becomes $i \sigma_y$, and together with complex conjugation transforms mass term into minus itself, i.e. the transformed Hamiltonian is
\begin{equation}
H'' = v_F \vec{\sigma} \vec{p} - \mu (\vec{r}) \sigma_z .
\end{equation}
The corresponding solution is
\begin{equation}
\Psi''  = \left[
  \begin{array}{c}
   \Psi_2^* \\
   - \Psi_1^*\\
  \end{array}
\right], \label{treigenstate}
\end{equation}
and represents the corresponding partner of (\ref{eigenstate}) under time reversal. In the graphene physics, the two states, (\ref{eigenstate}) and (\ref{treigenstate}), will correspond the fermionic states belonging to two different valleys - energy minima in $\vec{k}$ space.

We introduced the Dirac equation with a mass term to model the confining potential i.e. a monotonously increasing function of radius $r$ of the dot that confines fermions. This choice makes possible to fix and resolve straightforwardly the boundary condition as shown in Ref. \onlinecite{bm} in the case of an infinite mass potential. The boundary condition requires that the current along the radius at the circular boundary is zero, and the eigenstates have energy values that are real (infinite life times).

On the other hand if we consider the electrostatic confinement, $V(r)$, so that
\begin{equation}
H = v_F \vec{\sigma} \vec{p} + V(r),
\label{vc}
\end{equation}
in this case a hard-wall construction does not confine electrons which is a manifestation of the Klein paradox. For a finite  step-like electrostatic potential we have quasi-bound states as shown in Ref. \onlinecite{mp}. We note that at $E = 0$, bound states are possible for special potentials, as shown in Ref. \onlinecite{b0}. The structure of the quasi-bound states is very similar to the one that we get by using the infinite mass confinement as can be seen in Ref. \onlinecite{leeetal}, where the electrostatic confinement was considered both experimentally and theoretically by calculating spectra in a parabolic electrostatic confinement. The lowest energy state, with the maximum probability peaked at the origin, corresponds to the one in the infinite mass confinement with the same feature (and the rest of low-lying spectrum has a similar correspondence in terms of the maxima of probability) and we expect that at least qualitatively our conclusions will be valid also for the experimental set-up of Ref. \onlinecite{leeetal} and, in general, in the case of electrostatic confinement.

We proceed by reviewing the infinite mass confinement solutions (Ref. \onlinecite{pe}). The decoupling of the radial and angle part leads to the following form of the general solution with energy $E > 0$ in polar coordinates $r$ and $\phi$,
\begin{equation}
\Psi(r) = A \exp\{i m \phi\} \left[
  \begin{array}{c}
  J_m(k r) \\
   i \exp\{i  \phi\}  J_{m + 1}(k r)\\
  \end{array}
\right], \label{solutions}
\end{equation}
where the Bessel functions, $ J_m(k r)$, of the first kind, $ m = 0, \pm 1, \pm 2, \ldots$, $k = E /(\hbar v_F)$, and A is a normalization constant. The infinite mass boundary condition (Ref. \onlinecite{bm}) implies at $R$, the radius of the dot,
\begin{equation}
J_m(k_a R) = \tau  J_{m + 1}(k_a R),
\label{eqeigen}
\end{equation}
where $ \tau = \pm 1$ is the valley index and $a$ in $k_a$ denotes a set of indexes, $a = (n, m, \tau)$, that specify the eigenvalue $E_a = E_{(n, m, \tau)}$. Here $n$ enumerates discrete energy levels, $ n = 0, 1, 2, \ldots$ starting from the lowest one, with $m$ - orbital angular momentum quantum number and $\tau$ index.  We see that the degeneracy between the different valley solutions is described by the following relation,
\begin{equation}
E_{(n, m, \tau)} = E_{(n, - m - 1, - \tau)}.
\end{equation}

We will denote by $\Phi_a^{(+)}$ the normalized positive energy solutions,
\begin{equation}
\Phi_{a}^{(+)} (r) = A_a \exp\{i m \phi\} \left[
  \begin{array}{c}
    J_m(k_a r) \\
   i \exp\{i  \phi\} J_{m + 1}(k_a r)\\
  \end{array}
\right], \label{realsolutions}
\end{equation}
where $ k_a = E_a /(\hbar v_F) $, and the normalization factor is,
\begin{equation}
A_a = [ \pi (  J_m^2 -  J_{m - 1}  J_{m + 1} +  J_{m + 1}^2 -   J_m  J_{m + 2}]^{\frac{1}{2}},
\end{equation}
with $ J_m \equiv  J_m (E_a /\Delta_0) $ and $\Delta_0 = (\hbar v_F)/R$.

To introduce the Coulomb interaction  into the problem we consider the full electron second-quantized field operator,
\begin{equation}
\Psi(r) = \sum_a \Phi_{a}^{(+)} (r) c_a + \sum_a \Phi_{a}^{(-)} (r) d_a^\dagger ,
\label{defpsi}
\end{equation}
where we introduced  $\Phi_{a}^{(-)} (r)$, the negative energy solutions, which according to our previous discussion, must be of the following form,
\begin{equation}
\Phi_{a}^{(-)} (r) = A_a \exp\{- i m \phi\} \left[
  \begin{array}{c}
   - i \exp\{- i  \phi\}  J_{m + 1}(k_a r) \\
   J_m(k_a r) \\
  \end{array}
\right]. \label{negativesolutions}
\end{equation}
In Eq. (\ref{defpsi}), $c_a$ denotes the particle annihilation operator for level $a$, and $d_a^\dagger$ denotes the hole creation operator for the same level. The corresponding particle  creation operator is $c_a^\dagger$, and the hole annihilation operator is $d_a$.

We can decompose the full electron operator as  $\Psi(r) = \sum_{\tau, s} \Psi_{\tau, s}(r)$, where we introduced an additional flavor index $s ; s = 1, \ldots, N_f$; where $N_f$ denotes the number of flavors.

The Hamiltonian can be expressed as
\begin{equation}
H = H_K + H_I ,
\end{equation}
with the kinetic part,
\begin{equation}
H_K = \sum_a E_a c_a^\dagger c_a + \sum_a E_a d_a^\dagger d_a ,
\end{equation}
and the interaction term,
\begin{eqnarray}
H_I = \frac{\hbar v_F \alpha}{2} \sum_{\tau, s, \tau', s'} \int \frac{d \vec{r} d \vec{r}'}{|\vec{r} - \vec{r}'|} \times  \;\;\;\;\;\;\;\;\;\;\;\;\;\;\;\;\; \nonumber \\
 : \Psi_{\tau s}^{\dagger} (\vec{r}) \Psi_{\tau' s'}^{\dagger} (\vec{r}') \Psi_{\tau' s'} (\vec{r}') \Psi_{\tau s} (\vec{r}):,
\end{eqnarray}
where colons denote normal ordering, and $\alpha$ is the dimensionless parameter, $ \alpha = e^2 /(\hbar k v_F)$, which was introduced and discussed in the introduction section.

In Fig. 1 we can see the part of the spectrum that describes positive energy solutions, calculated from Eq. (\ref{eqeigen}) with $\tau = + 1 .$ The energy of the lowest energy state can be found to be $E_{(n=0, m=0, \tau=+1)} \equiv  E_{00} = 1.44 \Delta_0$. On the other hand we may estimate
that the Coulomb energy gained by the presence of an exciton, with a particle and a hole at distance (uncertainty) of the order of $R$, as $\alpha \Delta_0$. Thus if we want to study critical behavior and the process in which the first exciton is created under the steady increase of $\alpha$, from zero, we may concentrate on the lowest lying state, and its negative energy counterpart, and in that small subspace understand the influence of the valley and flavor degrees of freedom. The price we pay is certainly the absence of the higher energy physics, but we will be able to track down, in this small subspace, the processes that are relevant for the excitonic physics, which do not conserve separately particle and hole number, but only their difference.

Thus the subspace that we consider is made of two positive energy $ E = E_{00} > 0$ eigenstates:
\begin{equation}
\Phi_{(n=0,m=0,\tau=+1)}^{(+)} (r) = A_{00}  \left[
  \begin{array}{c}
    J_0(k_0 r) \\
   i \exp\{i  \phi\}  J_{1}(k_0 r)\\
  \end{array}
\right], \label{realsolution}
\end{equation}
and
\begin{equation}
\Phi_{(n=0,m=-1,\tau=-1)}^{(-)} (r) = A_{00}  \left[
  \begin{array}{c}
   -  \exp\{- i  \phi\}  J_{1}(k_0 r) \\
 i  J_0(k_0 r) \\
  \end{array}
\right], \label{negativesolution}
\end{equation}
and their negative energy counterparts. Here $k_0 = E_{00}/(\hbar v_F)$. Note that the states in (\ref{realsolution}) and (\ref{negativesolution}) are also eigenstates of the total angular momentum, $ J_z = - i \hbar \frac{\partial}{\partial \phi} + \frac{1}{2} \sigma_z$, and can be classified by its quantum number, $ M = m + 1/2$. Thus the two states in (\ref{realsolution}) and (\ref{negativesolution}), related by time reversal operation, are associated with $M = 1/2$ and $M = - 1/2$ quantum number respectively.
\newcounter{mathematicapage}

\begin{figure}[hbtp]

\centering
\begin{tikzpicture}
\begin{axis}[
	xlabel={$m$},
	ylabel={$\Delta E [(\hbar v_F)/R]$},
	height=7cm,
		width=8cm,
		];
\addplot [cyan, only marks,mark=+,mark size= 2pt, color=black] coordinates
{(0,1.3) (0,4.6) (0,7.9)(0,12) (1,2.6)(1,6.2)(1,9.5)(1,12.5)(2,3.8)(2,7.1)(2,10.9)(2,14)(3,4.9)(3,8.8)(3,12.3)(3,15.3)(4,6)(4,9.9)(4,13.4)(4,16.8)(-1,3.2)(-1,6.4)(-1,9.5)(-1,12.6)(-2,4.7)(-2,7.7)(-2,11.1)(-2,14.2)(-3,5.8)(-3,9.2)(-3,12.3)(-3,15.6)(-4,7)(-4,10.3)(-4,13.8)(-4,16.9)};
\end{axis}
\end{tikzpicture}

\caption{The energy spectrum  for the noninteracting case. Here m represents the quantum number of the orbital angular momentum.}
\end{figure}
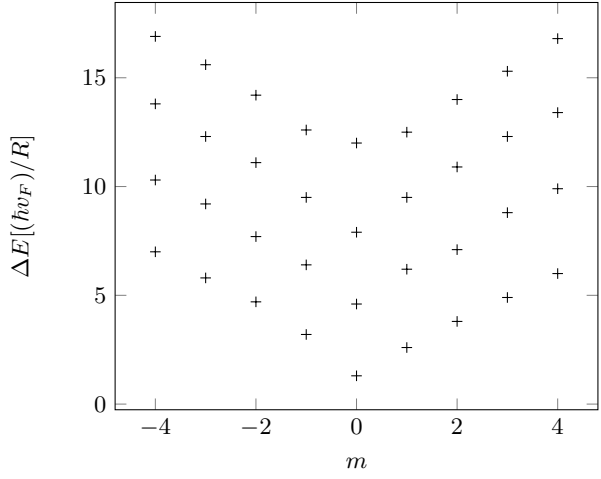

In the reduced, small subspace we denote the relevant  second-quantized creation and annihilation operators by $  c^\dagger_{\tau, s}, d^\dagger_{\tau, s}$, and
$  c_{\tau, s}, d_{\tau, s}$, where $ \tau = \pm 1$, and $ s = 1, \ldots, N_f$.
$c$'s stand for particles and $d$'s for holes.

 Using the solutions of the Dirac quantum dot we can express the electric (charge) density of the system as
\begin{eqnarray}
\rho (r) &=& \sum_{s,\tau} I_1 (r) c^\dagger_{\tau, s} c_{\tau, s} + \sum_{s,\tau} I_1 (r) d_{\tau, s} d^\dagger_{\tau, s} + \nonumber \\
&& \{ \sum_{s,\tau} I_2 (r) (-i) e^{-i \tau \phi} c^\dagger_{\tau, s} d^\dagger_{\tau, s} + h.c. \},
\end{eqnarray}
with
\begin{equation}
I_1 (r) = A_{00}^2 [  J_{0}^{2}(k_0 r) +  J_{0}^{2}(k_0 r)],
\end{equation}
and
\begin{equation}
I_2 (r) = A_{00}^2 \; 2 \;  J_{0}(k_0 r)  J_{1}(k_0 r).
\end{equation}

We define
\begin{equation}
\rho_d = \sum_{s,\tau}  c^\dagger_{\tau, s} c_{\tau, s} + \sum_{s,\tau}  d_{\tau, s} d^\dagger_{\tau, s}
\end{equation}
Then the normal ordered expression,
\begin{equation}
: \rho (r) \rho (r'):
\end{equation}
enters the formula for the Hamiltonian that describes Coulomb interaction, which final expression is
\begin{eqnarray}
{\cal H} & = & V_1 : \rho_d \rho_d : \nonumber \\
         & & - V_2 \sum_{\tau, s,s'} \{ c^\dagger_{\tau, s} d^\dagger_{\tau, s}  c^\dagger_{-\tau, s'} d^\dagger_{-\tau, s'} + h.c. \} \nonumber \\
         &&  + 2 \; V_2 \sum_{\tau, s,s'}  c^\dagger_{\tau, s} d^\dagger_{\tau, s} d_{\tau, s'} c_{\tau, s'},
\end{eqnarray}
where $V_1, V_2$ are the coefficients obtained by spatial integration over the Coulomb potential.

We define operators
\begin{equation}
B^\dagger_{\tau, s} = c^\dagger_{\tau, s} d^\dagger_{\tau, s},
\end{equation}
so that we can rewrite the interacting part of the Hamiltonian as
\begin{eqnarray}
{\cal H} & = & V_1 : \rho_d \rho_d : \nonumber \\
         & & - V_2 \sum_{\tau, s,s'} \{ B^\dagger_{\tau, s}  B^\dagger_{-\tau, s'} + h.c. \} \nonumber \\
         &&  + 2 \; V_2 \sum_{\tau, s,s'}  B^\dagger_{\tau, s} B_{\tau, s'}.
\end{eqnarray}
We also define
\begin{equation}
B_\tau = \sum_s B_{\tau, s} = \sum_s d_{\tau, s} c_{\tau, s} \label{Bdefined}
\end{equation}
for which we have the following commutation relations,
\begin{eqnarray}
[ B_\tau, B_\tau^\dagger ] & = & \sum_s ( - c^\dagger_{\tau, s} c_{\tau, s} +   d_{\tau, s} d^\dagger_{\tau, s} ) \nonumber \\
                           & = & N_f -  \hat{n}_{\tau}.
\end{eqnarray}
We can easily conclude that we can express the Hamiltonian with the help of $B_{\tau}$' s and that the following states,
\begin{eqnarray}
|n, N_f, n \leq N_f \rangle = \;\;\;\;\;\;\;\;\;\;\;\;\;\;\;\;\; \nonumber \\
\frac{1}{ (n!)^2 \left(
  \begin{array}{c}
   N_f \\
   n \\
  \end{array}
\right)}                 ( B^\dagger_\tau )^n ( B^\dagger_{-\tau} )^n |0\rangle , \label{SS0}
\end{eqnarray}
where $ n = 0,1, \ldots , N_f$, make an invariant subspace of the Hamiltonian. In Eq. (\ref{SS0}) we used parenthesis to denote the binomial coefficient,
\begin{equation}
\left( \begin{array}{c}
   N_f \\
   n \\
  \end{array}\right) = \frac{N_f !}{(N_f - n)! \; n!}.
\end{equation}
The $n = 0$ state is the vacuum state.

If we denote by $\delta {\cal H}_1$,
\begin{equation}
\delta {\cal H}_1 = V_1 : \rho_d \rho_d : ,
\end{equation}
we have
\begin{equation}
\langle n, N_f | \delta {\cal H}_1 |n, N_f \rangle = - 4 V_1 n .
\end{equation}
If we denote by $\delta {\cal H}_2$,
\begin{equation}
\delta {\cal H}_2 = - V_2 \sum_{\tau, s,s'}  B^\dagger_{\tau, s} B^\dagger_{-\tau, s'} = -  V_2 \sum_{\tau}  B^\dagger_{\tau} B^\dagger_{-\tau},
\end{equation}
we have
\begin{equation}
\langle n + 1, N_f | \delta {\cal H}_2 |n, N_f \rangle = - 2 V_2 (n + 1) (N_f - n).
\end{equation}
If we denote by $\delta {\cal H}_3$,
\begin{equation}
\delta {\cal H}_3 =  2 V_2 \sum_{\tau, s,s'}  B^\dagger_{\tau, s} B_{\tau, s'} = 2 V_2 \sum_{\tau}  B^\dagger_{\tau} B_{\tau},
\end{equation}
we have
\begin{equation}
\langle n, N_f | \delta {\cal H}_3 |n, N_f \rangle = 4 n V_2 (N_f - n + 1).
\end{equation}

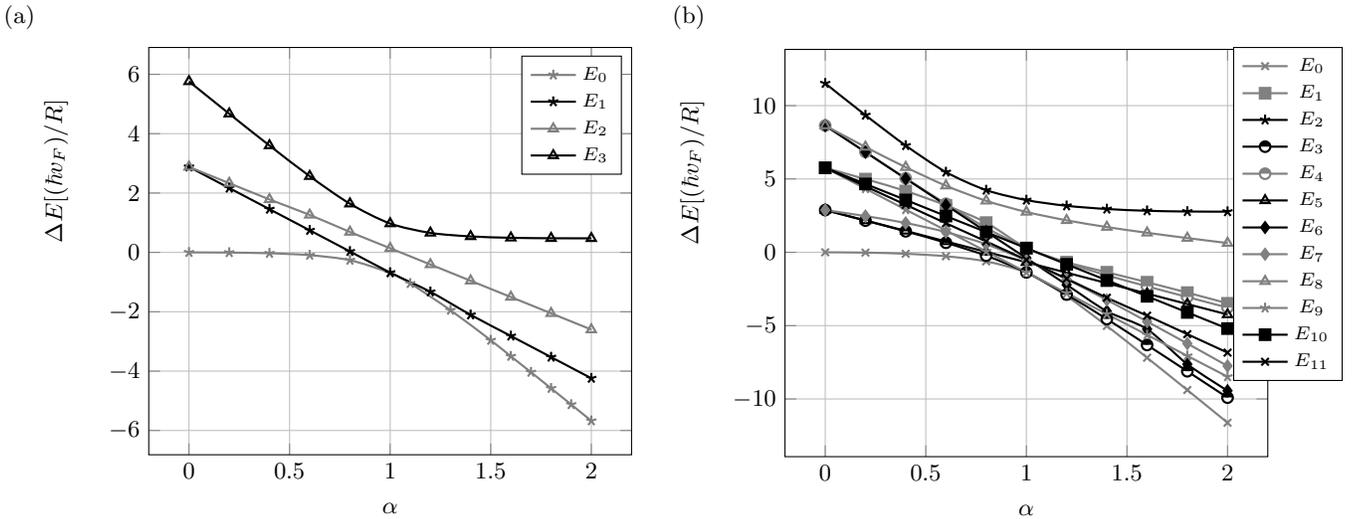
\begin{figure*}[t]
\centering
\begin{subfigure}[t]{0.49\textwidth}
\caption{}
\begin{tikzpicture}
\begin{axis}[
	xlabel={$\alpha$},
	ylabel={$\Delta E [(\hbar v_F)/R]$},
	height=7cm,
		width=8cm,
		grid=major,
        right
,legend style={font=\fontsize{7}{8}\selectfont}]
\addplot [smooth, thick, color =gray, mark =star,mark size= 2pt]
  plot coordinates {
 (0, 0) (0.2,-0.006) (0.4,-0.03)
 (0.6,-0.094)  (0.8,-0.263)  (1.0,-0.692)
 (1.1,-1.046) (1.3,-1.948) (1.5,-2.96) (1.6,-3.5)
 (1.7,-4.037) (1.8,-4.58) (1.9,-5.126) (2,-5.674)};
\addplot [smooth, thick, color = black, mark =star,mark size= 2pt]
  plot coordinates {
(0, 2.88)  (0.2,2.168) (0.4,1.456)
 (0.6,0.744) (0.8,0.032) (1.0,-0.68)
 (1.2,-1.329) (1.4,-2.104) (1.6,-2.816)
(1.8,-3.528) (2,-4.24)};
\addplot [smooth, thick, color =gray, mark =triangle,mark size= 2pt]
  plot coordinates {
 (0, 2.88) (0.2,2.332) (0.4,1.784)
 (0.6,1.263) (0.8,0.688) (1.0,0.14)
 (1.2,-0.408) (1.4,-0.956) (1.6,-1.504)
 (1.8,-2.052) (2,-2.6)};
\addplot [smooth, thick, color =black, mark =triangle,mark size= 2pt]
  plot coordinates {
 (0, 5.76) (0.2,4.67) (0.4,3.6)
(0.6,2.566)  (0.8,1.639) (1.0,0.972)
  (1.2,0.657) (1.4,0.538) (1.6,0.492)
 (1.8,0.476) (2,0.474)};
\legend{$E_{0}$,$E_{1}$,$E_{2}$,$E_{3}$}
\end{axis}
\end{tikzpicture}
\end{subfigure}
\begin{subfigure}[t]{0.49\textwidth}
\caption{}
\begin{tikzpicture}
\begin{axis}[xlabel={$\alpha$},
	ylabel={$\Delta E [(\hbar v_F)/R]$},
	height=7cm,
		width=8cm,
		grid=major,
		right
,legend style={at={(axis cs:2.3,14)},anchor=south west,anchor=north,font=\fontsize{7}{8}\selectfont}]
]

\addplot [smooth, thick, color =gray, mark =x,mark size= 2pt]
  plot coordinates{
	(0,0) (0.2,-0.022) (0.4,-0.1)
    (0.6,-0.27 ) (0.8,-0.62 ) (1.0,-1.39 )
    (1.2,-2.99) (1.4,-5.02) (1.6,-7.19 )
    (1.8,-9.39 ) (2,-11.62)};
\addplot [smooth, thick, color = gray, mark =square*,mark size= 2pt]
  plot coordinates{
	(0,5.76) (0.2,4.99) (0.4,4.18)
    (0.6,3.2 ) (0.8,2.025) (1.0,0.3 )
    (1.2,-0.67) (1.4,-1.36) (1.6,-2.04 )
    (1.8,-2.75 ) (2,-3.48)};
\addplot [smooth, thick, color =black, mark =star,mark size= 2pt]
  plot coordinates  {
	(0,11.52) (0.2,9.35) (0.4,7.28)
    (0.6,5.47 ) (0.8,4.24) (1.0,3.56 )
    (1.2,3.17) (1.4,2.95) (1.6,2.83 )
    (1.8,2.78 ) (2,2.77)};

\addplot [smooth, thick, color =black, mark =halfcircle*, mark size= 2pt]
 plot coordinates{
	(0,2.88) (0.2,2.16) (0.4,1.43)
    (0.6,0.65 ) (0.8,-0.23) (1.0,-1.37 )
    (1.2,-2.88) (1.4,-4.55) (1.6,-6.32 )
    (1.8,-8.11 ) (2,-9.91)};
\addplot [smooth, thick, color =gray, mark =halfcircle*,mark size= 2pt]
  plot coordinates  {
	(0,8.64) (0.2,6.84) (0.4,5.05)
    (0.6,3.31 ) (0.8,1.67) (1.0,0.29 )
    (1.2,-0.73) (1.4,-1.57) (1.6,-2.32 )
    (1.8,-3.05 ) (2,-3.77)};
 \addplot [smooth, thick, color =black, mark =triangle,mark size= 2pt]
  plot coordinates  {
	(0,2.88) (0.2,2.17) (0.4,1.46)
    (0.6,0.74) (0.8,0.03) (1.0,-0.68 )
    (1.2,-1.39) (1.4,-2.104) (1.6,-2.816 )
    (1.8,-3.53 ) (2,-4.24)};
\addplot [smooth, thick, color = black, mark = diamond*,mark size= 2pt]
  plot coordinates{
	(0,8.64) (0.2,6.83) (0.4,5.02)
    (0.6,3.22 ) (0.8,1.41) (1.0,-0.4 )
    (1.2,-2.21) (1.4,-4.02) (1.6,-5.24 )
    (1.8,-7.63 ) (2,-9.44)};
\addplot [smooth, thick, color =gray, mark = diamond*,mark size= 2pt]
  plot coordinates{
	(0,2.88) (0.2,2.47) (0.4,2)
    (0.6,1.38) (0.8,0.55) (1.0,-0.55)
    (1.2,-1.84) (1.4,-3.25) (1.6,-4.72 )
   (1.8,-6.22 ) (2,-7.75)};
\addplot [smooth, thick, color =gray, mark =triangle,mark size= 2pt ]
  plot coordinates{
	(0,8.64) (0.2,7.18) (0.4,5.8)
    (0.6,4.54 ) (0.8,3.51) (1.0,2.75 )
    (1.2,2.18) (1.4,1.72) (1.6,1.33 )
    (1.8,0.97 ) (2,0.63)};

\addplot [smooth, thick, color = gray, mark =star,mark size= 2pt]
  plot coordinates{
	(0,5.76) (0.2,4.34) (0.4,2.91)
    (0.6,1.49 ) (0.8,0.06) (1.0,-1.36)
    (1.2,-2.78) (1.4,-4.21) (1.6,-5.63 )
    (1.8,-7.06 ) (2,-8.48)};
\addplot [smooth, thick, color = black, mark =square*,mark size= 2pt]
  plot coordinates{
	(0,5.76) (0.2,4.66) (0.4,3.57)
    (0.6,2.47 ) (0.8,1.38) (1.0,0.28 )
    (1.2,-0.82) (1.4,-1.91) (1.6,-3 )
    (1.8,-4.1 ) (2,-5.2)};
\addplot [smooth, thick, color =black, mark = x,mark size= 2pt]
  plot coordinates{
	(0,5.76) (0.2,4.5) (0.4,3.24)
    (0.6,1.98 ) (0.8,0.74) (1.0,-0.52 )
    (1.2,-1.8) (1.4,-3.1) (1.6,-4.32 )
    (1.8,-5.58 ) (2,-6.84)};
\legend{$E_{0}$,$E_{1}$,$E_{2}$,$E_{3}$,$E_{4}$,$E_{5}$,$E_{6}$,$E_{7}$,$E_{8}$, $E_{9}$,$E_{10}$,$E_{11}$}
\end{axis}
\end{tikzpicture}

\end{subfigure}
\caption{The eigenspectrum for (a) $N_f = 1$, and (b) $N_f = 2$. See
Appendix A for explicit formulas of the plotted energies, $\{ E_i ; i =
0, 1, 2, 3\}$ in (a). See main text and Appendix B for explicit formulas of the
following energies, $\{ E_i ; i = 3, 4,\ldots, 11\}$ in (b). $E_0, E_1,$
and $E_2$ in (b) are numerically obtained.
}
\end{figure*}

\begin{figure*}[t]
\centering
\begin{subfigure}[t]{0.49\textwidth}
\caption{}
\begin{tikzpicture}
\begin{axis}[
	xlabel={$\alpha$},
	ylabel={$P(\alpha)$},
	height=7cm,
		width=8cm,
		grid=major,
		right
,legend style={font=\fontsize{7}{8}\selectfont,at={(axis cs:0,0.3)},anchor=south west}
]
\addplot [smooth, thick, color =black, mark =triangle,mark size= 2pt]
  plot coordinates{
	(0,1) (0.2,0.999 ) (0.4,0.992 )
    (0.6,0.965) (0.8,0.862) (1.0,0.584 )
   (1.2,0.309) (1.4,0.18) (1.6,0.123)
    (1.8,0.094 ) (2,0.077)};
\addplot [smooth, thick, color =gray, mark =halfcircle*,mark size= 2pt]
  plot coordinates{
	(0,0) (0.2,0.001) (0.4,0.008 )
	(0.6,0.036 ) (0.8,0.138) (1.0,0.416 )
	(1.2,0.692 ) (1.4,0.82) (1.6,0.877 )
	(1.8,0.906 ) (2,0.923)};
\legend{$P_{1}$,$P_{2}$}
\end{axis}
\end{tikzpicture}
\end{subfigure}
\begin{subfigure}[t]{0.49\textwidth}
\caption{}
\centering
\begin{tikzpicture}
\begin{axis}[
	xlabel={$\alpha$},	ylabel={$P(\alpha)$},
	height=7cm,		width=8cm,		grid=major,
		right
,legend style={font=\fontsize{7}{8}\selectfont,at={(axis cs:0,0.3)},anchor=south west}]
\addplot [smooth, thick, color = black, mark =star,mark size= 2pt]
  plot coordinates{
	(0,1) (0.2,0.996 ) (0.4,0.98)
    (0.6,0.93 ) (0.8,0.82 )(0.9,0.71) (1.0,0.53 )
    (1.1,0.32) (1.2,0.17)(1.3,0.093) (1.4,0.057) (1.6,0.028)
    (1.8,0.017)(1.9,0.014) (2,0.012)};
\addplot [smooth, thick, color =gray, mark =triangle,mark size= 2pt]
  plot coordinates{
	(0,0) (0.2,0.001 ) (0.4,0.02)
    (0.6,0.07 ) (0.8,0.16 )(0.9,0.26) (1.0,0.32 )
    (1.1,0.33)(1.2,0.28)(1.3,0.24) (1.4,0.21) (1.6,0.16)
    (1.8,0.14)(1.9,0.14) (2,0.137)};
 \addplot [smooth, thick, color = black, mark =halfcircle*,mark size= 2pt]
  plot coordinates {
	(0,0) (0.2,0.000005 ) (0.4,0.0002)
    (0.6,0.003 ) (0.8,0.026 )(0.9,0.083) (1.0,0.23 )
    (1.1,0.43 )(1.2,0.59)(1.3,0.68) (1.4,0.74) (1.6,0.81)
    (1.8,0.85) (1.9,0.8596) (2,0.869)};
\legend{$P_{1}$,$P_{2}$,$P_{3}$}
\end{axis}
\end{tikzpicture}
\end{subfigure}
\caption{Exciton number and polarization  of the ground state for (a)
$N_f = 1$ (the participations of the states with 0 and 2 excitons), and
  (b) $N_f = 2$ (the participations of the states with 0, 2, and 4
excitons).}
\end{figure*}
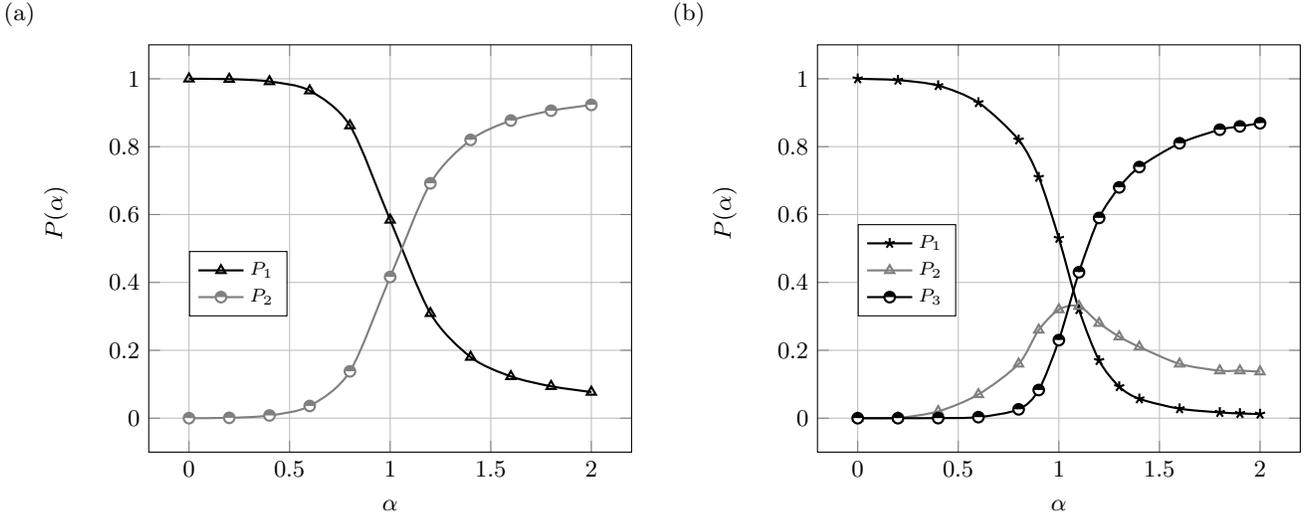

Although very useful for the determination of the ground state energy (due to the reduction to smaller space) the subspace does not give complete picture of the spectrum. In Fig. 2  we can view the complete spectrum when
$N_f = 1$ and $N_f = 2$ respectively. In the Appendix the complete classification of eigenvectors and eigenstates is given for both cases.

From the solutions of the complete problem  in the $ N_f = 1$ case and  in the $ N_f = 2$ case, we can conclude that the
ground state in a crossover manner evolves from the vacuum to a highly polarized
state with increase of the strength of the Coulomb interaction measured by the
$\alpha$ coefficient.

In Fig. 2b, we can view the graphene, $ N_f = 2$, case. The eigenvalues  $E_0, E_1,$ and $E_2$ are numerically obtained, while the rest of the energies can be expressed as:
$E_{3(-)/4(+)}=2\left(2 E_{00} - 2 V_{1} + V_{2} \mp \sqrt{(E_{00} - V_{1} + V_{2})^{2}+ V_{2}^{2}}\right)$,   $E_{5}=2\left(E_{00} - V_{1}\right)$,
$E_{6}=2\left(3 E_{00} - 3 V_{1} + 2 V_{2}\right)$,
$E_{7(-)/8(+)}=2\left(2 E_{00} - 2 V_{1} + 3 V_{2} \mp \sqrt{(E_{00} - V_{1} + V_{2})^{2} + 4 V_{2}^{2}}\right)$, $E_{9}=4\left(E_{00} - V_{1}\right)$,
$E_{10}=4\left(E_{00} - V_{1} + V_{2}\right)$,
$E_{11}=2\left(2 E_{00} - 2 V_{1} + V_{2}\right)$.
In these expressions, $E_{00}$ represents the kinetic term.

The analysis of the  $ N_f = 2$ makes clear that the state(s) which energy is
the closest to the ground state energy at arbitrary $\alpha$ comes from the invariant subspace made by the following states
\begin{eqnarray}
|n, N_f, \sigma , \sigma', n \leq N_f - 1 \rangle = \;\;\;\;\;\;\;\;\;\;\;\;\;\;\;\;\; \nonumber \\
\frac{1}{ (n!)^2 \left(
  \begin{array}{c}
   N_f - 1 \\
   n \\
  \end{array}
\right)}                 ( B^\dagger_\tau )^n ( B^\dagger_{-\tau} )^n c^\dagger_{\tau, \sigma} d^\dagger_{-\tau, \sigma'} |0\rangle ,
\label{SS1}
\end{eqnarray}
where, $\sigma$ and $\sigma'$ represent any two flavors (that are allowed also to be the same);
$ \sigma , \sigma' \in \{1,2, \ldots, N_f\}$.

In each special subspace defined by fixed $ \sigma , \sigma'$, we have

\begin{equation}
\langle n, N_f, \sigma , \sigma' | \delta {\cal H}_1 |n, N_f, \sigma , \sigma' \rangle = - (4 n + 2) V_1  ,
\end{equation}
and
\begin{equation}
\langle n + 1, N_f, \sigma , \sigma' | \delta {\cal H}_2 |n, N_f, \sigma , \sigma' \rangle = - 2 V_2 (n + 1) (N_f - 1 - n),
\end{equation}
and
\begin{equation}
\langle n, N_f, \sigma , \sigma' | \delta {\cal H}_3 |n, N_f, \sigma , \sigma' \rangle = 4 n V_2 (N_f - n).
\end{equation}

If we compare the diagonal energy of the state $ |n = N_f \rangle $ of the subspace in (\ref{SS0}), which we denote by $E_a$, and the diagonal energy of the state $ |n = N_f -1, \sigma, \sigma' \rangle $ of the subspace in (\ref{SS1}),
which we denote by $E_b$, we see that
\begin{equation}
E_b - E_a = 2 (V_1 - 2 V_2 - E_{00}),
\end{equation}
where by $E_{00}$ we denoted the lowest energy level of the kinetic part. For
$ V_1 > 2 V_2 $, and for large enough $\alpha$, we expect that the correlated state connected with vacuum,
with even number of excitons, and the large participation of the state with the maximum number of excitons  equal to $2 N_f$, will always have lower energy
than the correlated state with odd number of excitons, and the large participation of the state with the number of excitons equal to $2 N_f -1 $. This is indeed the case when $N_f = 1 $ and $N_f = 2 $.

\begin{figure*}[!t]
\centering
\begin{subfigure}[t]{0.49\textwidth}
\caption{}
\begin{tikzpicture}
\begin{axis}[
	xlabel={$\alpha$},
	ylabel={$\Delta E [(\hbar v_F)/R]$},
	height=7cm,
		width=8cm,
		grid=major,
		right
,legend pos=south west,legend style={font=\fontsize{7}{8}\selectfont}]
\addplot [smooth, thick, color =black, mark =triangle,mark size= 2pt]
  plot coordinates {
	(0,0) (0.2,-0.022 ) (0.4,-0.1 )
    (0.6,-0.27 ) (0.8,-0.62 ) (1.0,-1.39 )
    (1.2,-2.99) (1.4,-5.031) (1.6,-7.19)
    (1.8,-9.4 ) (2,-11.63 )};
\addplot [smooth, thick, color = gray, mark =halfcircle*,mark size= 2pt]
  plot coordinates {
	(0,2.88) (0.2,2.16 ) (0.4,1.43 )
	(0.6,0.65 ) (0.8,-0.23) (1.0,-1.37 )
	(1.2,-2.87 ) (1.4,-4.55 ) (1.6,-6.32 )
	(1.8,-8.11 ) (2,-9.91 )};
\legend{$E_{0}$,$E_{1}$}
\end{axis}
\end{tikzpicture}
\end{subfigure}
\begin{subfigure}[t]{0.49\textwidth}
\caption{}
\begin{tikzpicture}
\begin{axis}[
	xlabel={$\alpha$},
	ylabel={$\Delta E [(\hbar v_F)/R]$},
	height=7cm,
		width=8cm,
		grid=major,
		right
,legend pos=south west,legend style={font=\fontsize{7}{8}\selectfont}]
\addplot [smooth, thick, color =black, mark =triangle,mark size= 2pt]
  plot coordinates {
	(0,0) (0.2,-0.08 ) (0.4,-0.32 )
   (0.6,-0.75 ) (0.8,-1.45 ) (1.0,-2.81 )
   (1.2,-6.13) (1.4,-10.4) (1.6,-14.9 )
    (1.8,-19.35 ) (2,-23.89 )};
\addplot [smooth, thick, color =gray, mark =halfcircle*,mark size= 2pt]
 plot coordinates {
	(0,2.88) (0.2,2.12 ) (0.4,1.26 )
	(0.6,0.25 ) (0.8,-0.99 ) (1.0,-2.77 )
	(1.2,-5.94 ) (1.4,-9.79 ) (1.6,-13.8 )
	(1.8,-17.87 ) (2,-22 )};
\legend{$E_{0}$,$E_{1}$}
\end{axis}

\end{tikzpicture}
\end{subfigure}
\caption{The two lowest energies of the two invariant subspaces,
expressed by Eq.(\ref{SS0}) and Eq.(\ref{SS1}) for (a)  $N_f = 2$, and
(b) $N_f =4 $. }
\end{figure*}
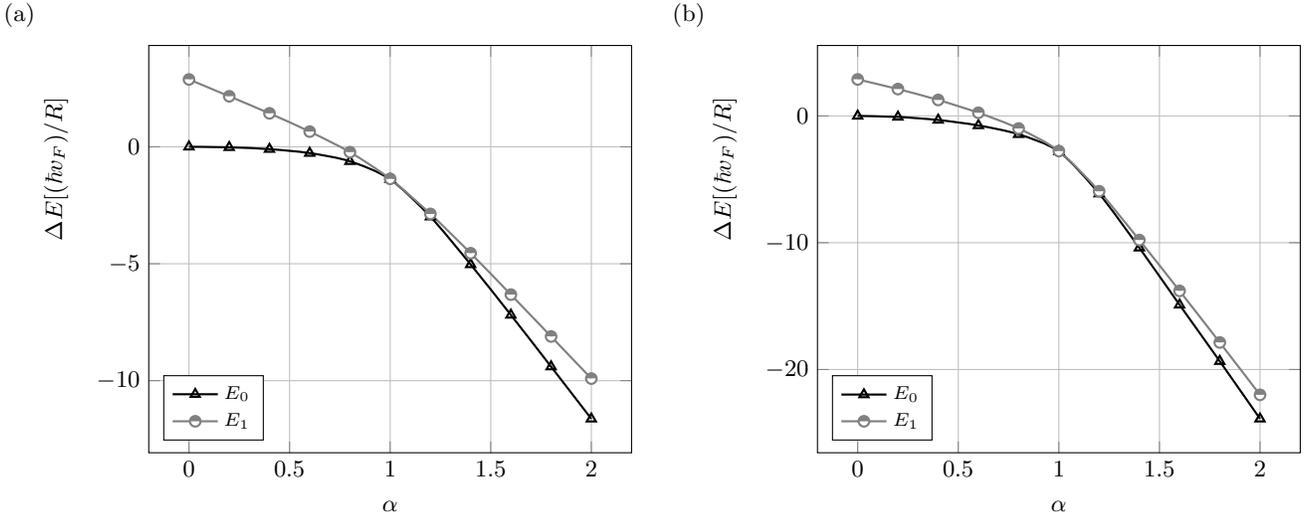

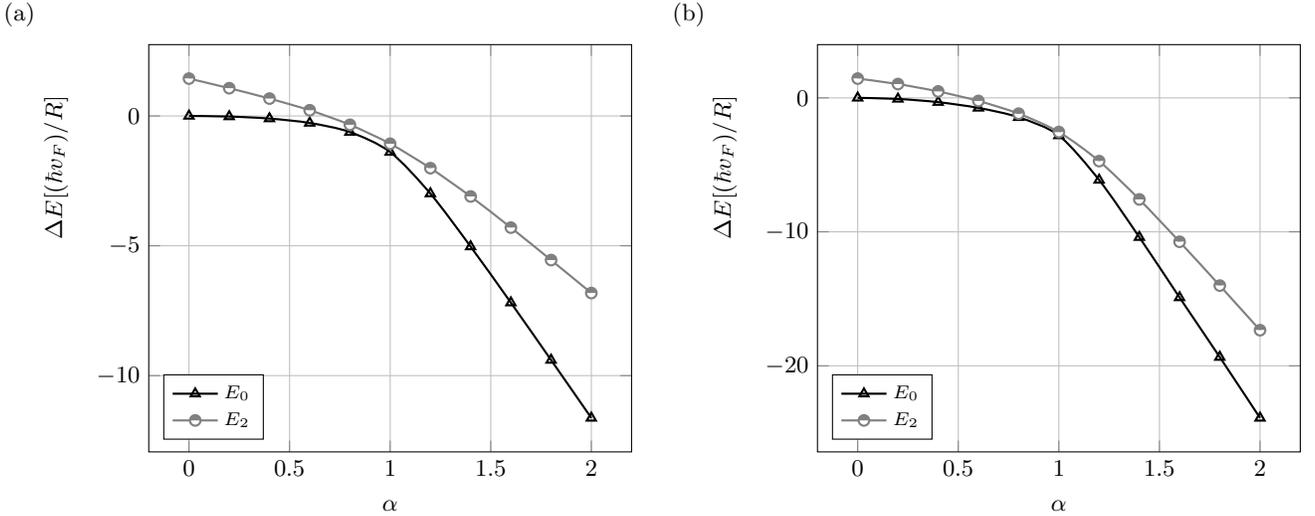
\begin{figure*}[t]
\centering
\begin{subfigure}[t]{0.49\textwidth}
\caption{}
\centering
\begin{tikzpicture}
\begin{axis}[
	xlabel={$\alpha$},
	ylabel={$\Delta E [(\hbar v_F)/R]$},
	height=7cm,
		width=8cm,
		grid=major,
		right
,legend pos=south west,legend style={font=\fontsize{7}{8}\selectfont}]
\addplot [smooth, thick, color =black, mark =triangle,mark size= 2pt]
  plot coordinates{(0,0) (0.2,-0.022 ) (0.4,-0.1 )
    (0.6,-0.27 ) (0.8,-0.62 ) (1.0,-1.39 )
    (1.2,-2.99) (1.4,-5.031) (1.6,-7.19)
    (1.8,-9.4 ) (2,-11.63 )};
     \addplot [smooth, thick, color =gray, mark =halfcircle*, mark size= 2pt]
  plot coordinates {
	(0,1.44) (0.2,1.07) (0.4,0.67 )
	(0.6,0.22 ) (0.8,-0.34) (1.0,-1.07 )
	(1.2,-2.01 ) (1.4,-3.1) (1.6,-4.3 )
	(1.8,-5.55 ) (2,-6.82 )};
\legend{$E_{0}$,$E_{2}$}
\end{axis}
\end{tikzpicture}
\end{subfigure}
\begin{subfigure}[t]{0.49\textwidth}
\caption{}
\centering
\begin{tikzpicture}
\begin{axis}[
	xlabel={$\alpha$},
	ylabel={$\Delta E [(\hbar v_F)/R]$},
	height=7cm,
		width=8cm,
		grid=major,
		right
,legend pos=south west,legend style={font=\fontsize{7}{8}\selectfont}]
\addplot [smooth, thick, color =black, mark =triangle,mark size= 2pt]
  plot coordinates{
	(0,0) (0.2,-0.08 ) (0.4,-0.32 )
    (0.6,-0.75 ) (0.8,-1.45 ) (1.0,-2.81 )
    (1.2,-6.13) (1.4,-10.4) (1.6,-14.9 )
   (1.8,-19.35 ) (2,-23.89 )};
\addplot [smooth, thick, color =gray, mark =halfcircle*,mark size= 2pt]
  plot coordinates {
	(0,1.44) (0.2,1.03) (0.4,0.48 )
	(0.6,-0.23 ) (0.8,-1.18) (1.0,-2.55 )
	(1.2,-4.72 ) (1.4,-7.58) (1.6,-10.74 )
	(1.8,-14.01 ) (2,-17.34 )};
\legend{$E_{0}$,$E_{2}$}
\end{axis}
\end{tikzpicture}
\end{subfigure}
\caption{The energy difference, as a function of $\alpha$, between the
lowest energy state in subspace Eq.(\ref{SS2}), and the one in subspace
described by Eq.(\ref{SS0}) i.e. the ground state for (a) $N_f = 2$, and
(b) $N_f = 4$.}
\end{figure*}

We notice that for $\alpha_c \approx 1.05$ i.e. when $E_{00} - V_1 + V_2 \approx 0$ all participations of the states in  (\ref{SS0}) of the invariant subspace in the ground state  are equal, and the lowest energy level is half-filled with excitons. In the cases with $ N_f = 1$ and $N_f = 2$, the participations with definite exciton numbers can be seen in Fig. 3, respectively. At the same time the lowest energy eigenstates that belong to the two invariant subspaces (expressed by Eq.(\ref{SS0}) and Eq.(\ref{SS1})) are degenerate - see Fig. 4. They represent states with different parities of the exciton number, in which the state with odd parity is characterized by the presence of a single intervalley exciton.  Notice that from the degenerate states from different subspaces described by Eq.(\ref{SS1}) we
can make superpositions that are invariant under the time reversal.

On the other hand we can consider a situation when an electron of valley $\tau$ and flavor index $\sigma$ is brought to the lowest level that we consider. The relevant, invariant subspace in that case is
\begin{eqnarray}
|n, N_f, \sigma , n \leq N_f - 1 \rangle = \;\;\;\;\;\;\;\;\;\;\;\;\;\;\;\;\; \nonumber \\
\frac{1}{ (n!)^2 \sqrt{\left(
  \begin{array}{c}
   N_f - 1 \\
   n \\
  \end{array}
\right)} \sqrt{\left(
  \begin{array}{c}
   N_f \\
   n \\
  \end{array}
\right)}}                 ( B^\dagger_\tau )^n ( B^\dagger_{-\tau} )^n c^\dagger_{\tau, \sigma} |0\rangle ,
\label{SS2}
\end{eqnarray}
In each special subspace defined by fixed $ \sigma $, we have

\begin{equation}
\langle n, N_f, \sigma | \delta {\cal H}_1 |n, N_f, \sigma  \rangle = - (4 n + 1) V_1  ,
\end{equation}
and
\begin{eqnarray}
\langle n + 1, N_f, \sigma  | \delta {\cal H}_2 |n, N_f, \sigma  \rangle = \nonumber \\
 - 2 V_2 (n + 1)\sqrt{(N_f - 1 - n)(N_f - n)},
\end{eqnarray}
and
\begin{equation}
\langle n, N_f, \sigma  | \delta {\cal H}_3 |n, N_f, \sigma  \rangle = 4 n V_2 (N_f - n) + 2 V_2 n.
\end{equation}
In Fig. 5 we can see that around $ \alpha \sim 1$ the energy of the added electron is nearly degenerate with the ground state energy. We find this behavior irrespective of the number of flavors.

\section{Discussion and Conclusions}

We presented a description of vacuum polarization in a circular Dirac quantum dot in two spatial dimensions assuming
$\alpha$ - the relative strength of the Coulomb interaction small enough to render an approximation with a single
electron (hole) lowest energy level relevant. Applying this approximation, we find that for $\alpha_c \approx 1.05$ the
lowest level is half-filled irrespective of the number of flavors that are present.

We also find that the lowest energy state can be described as an ``excitonic cloud" i.e. a superposition of excitonic states given by Eq. (\ref{SS0}) with (\ref{Bdefined}). The first level above can be described as an intervalley exciton, modified by the presence of other excitons (excitonic cloud). This state is always given by a superposition of states  described by Eq. (\ref{SS1}). These two levels, the ground and intervalley excitonic state, touch at  $\alpha_c
\approx 1.05$ when the average number of excitons is half the maximum occupancy of a single (particle, hole) level. The renormalized intervalley exciton is always below intravalley excitations due to the exchange effect, which is easily identified in the case with no flavors (only valley degree of freedom) in Fig. 2a. It is interesting to note that if we were to surround our dot, not with trivial but a topological insulator, in the infinite mass limit, in the case with only valley degree of freedom, there would be a real level crossing of the (unpolarized) vacuum state with the intervalley exciton state, degenerate with intravalley exciton state.

Although our approach is certainly relevant for $\alpha << 1$, due to the half-filling at $\alpha \sim 1$, we expect that the crossover behavior for $\alpha \lesssim 1$ that we detect, see Fig. 2, would remain if higher energy levels are included. Thus our results are reliable for $\alpha \lesssim 1$. We expect that the critical behavior that we described: level narrowing, and electron and exciton (near) resonances  will be present as we are approaching $\alpha_c$. That is in particular relevant for graphene on the $Si 0_2$ substrate with $ \alpha \simeq 0.79 $. Certainly, in the case of free standing  graphene,
$\alpha \simeq 2.16 $, at least one more level should be included and this is a subject for future investigations.

Our results are unaffected by screening because, at the neutrality point, with no finite density of charge, screening is suppressed in graphene. Ripples and disorder (which will induce puddles of particles and holes) will break the particle-hole symmetry, and hinder the creation of excitons, and this may blur the critical behavior we described.

It follows from our results that, due to the particle-hole symmetry, at the neutrality point, the first energy level in the particle spectrum will be pushed effectively downwards (less energy needed to bring an electron, see Fig. 5), and the first energy level in the hole spectrum will be pushed effectively upwards. This should be a consequence of the inclusion of electron-electron (i.e. Coulomb) interaction. Unfortunately, experimentally it is hard to realize a mass confinement,  in order to make a direct comparison with our results. Nevertheless, as we already discussed at the
beginning of Section II, electrostatic  confinements may serve as experimental set-ups that may show some of the behavior we find. Indeed, the basic prediction of the relative shift of the first (hole) energy level (with $\alpha \neq 0$, see Fig. 5a) may be relevant for the explanation of the experimental data in Ref. \onlinecite{guietal}.
Namely, considering the negative energy (hole) states in an anti-dot electrostatic confinement, it was noticed that there is a discrepancy between the experimental and calculated energy of the first energy level. As we described, the inclusion of the Coulomb interaction is important for the description of the lowest lying state(s) in graphene, and the discrepancy i.e. shift towards the neutrality point energy of the ``bare" - non-interacting level, may be an explanation for the observed discrepancy. It would be interesting to use different substrates to make anti-dots, in order to find out how shift depends on the effective $\alpha$ (which can be measured as shown in Ref. \onlinecite{guietal}). Our expectation is that, upon decreasing of $\alpha$, the observed discrepancy will be smaller.

It is interesting to note that for some special confining potentials for which $V({\bf r}) \sim |{\bf r}|$ in Eq. (\ref{vc}), we may map the (non-interacting) eigenvalue problem to one in an effective magnetic field [\onlinecite{r1,r2}].
The ensuing eigenstates include degenerate zero energy states, as well othe low-lying states that must be taken into account if the reduced space approximation discussed here is considered. A reduction was already considered in Ref. \onlinecite{r1}, not in the case of a single dot, but in the case of an array of quantum dots to find the dispersion of collective (``magneto-plasmon") states that propagate along the system. The potential, $V({\bf r}) \sim |{\bf r}|$, is special because of the non-analytic behavior at $ |{\bf r}| = 0$, in the real space, and does not represent a usual modeling of confinement that was considered in this work. A generalization to an array of dots, possibly in the presence of magnetic field, may be a direction for future work. This is motivated by the search for devices that will control and modify graphene transport properties.

\begin{acknowledgments}
We would like to thank E. Dobard\v zi\'c for help in the early stages of this work and I. Vasi\'c.
The work was supported by
the Ministry of Education, Science, and Technological
Development of the Republic of Serbia under project
ON171017.
\end{acknowledgments}

\appendix

\section{$N_f = 1$}

In this Appendix A we will present the complete solutions: eigenvectors and eigenvalues, when $N_f = 1$,
of our effective Hamiltonian for the circular graphene dot.

The subspace described by Eq. (\ref{SS0}) in this case consists of the vectors,
\begin{equation}
|0 \rangle , B_\tau^\dagger  B_{-\tau}^\dagger |0 \rangle \equiv c_+^\dagger d_+^\dagger c_-^\dagger d_-^\dagger |0 \rangle .
\end{equation}
The eigenproblem  in this invariant subspace is represented by the following matrix,
\begin{equation}
\left[
  \begin{array}{cc}
   0 & - 2 V_2 \\
   - 2 V_2 & 4 (E_{00} - V_1 + V_2) \\
  \end{array}
\right].
\end{equation}
The eigenvalues are $E_{0,3} = 2 (E_{00} - V_1 + V_2) \mp \sqrt{4 (E_{00} - V_1 + V_2)^2 + 4 V_2^2}$
with corresponding eigenvectors. The remaining vectors, $ B_\tau^\dagger |0 \rangle $, $\tau = \pm$, are eigenvectors with eigenvalue $ E_2 = 2 (E_{00} - V_1 + V_2)$, and $  c_\tau^\dagger d_{- \tau}^\dagger |0 \rangle $, $\tau = \pm$, are eigenvectors with eigenvalue $ E_1 = 2 (E_{00} - V_1)$. The resulting spectrum and its dependence on $\alpha$ can be seen in Fig. 2a.

\section{$N_f = 2$}

In this Appendix B we will present the complete solutions: eigenvectors and eigenvalues, when  $N_f = 2$,
of our effective Hamiltonian for the circular graphene dot.

The subspace described by Eq. (\ref{SS0}) in this case consists of the vectors,
\begin{equation}
|0 \rangle , \frac{1}{2} B_\tau^\dagger  B_{-\tau}^\dagger |0 \rangle , \frac{1}{4} (B_\tau^\dagger)^2  (B_{-\tau}^\dagger)^2 |0 \rangle .
\end{equation}
The eigenproblem  in this invariant subspace is represented by the following matrix,
\begin{equation}
\left[
  \begin{array}{ccc}
   0 & - 4 V_2 & 0 \\
   - 4 V_2 & 4 (E_{00} - V_1 + 2 V_2) & - 4 V_2 \\
   0 & - 4 V_2 & 8 (E_{00} - V_1 + V_2)  \\
  \end{array}
\right].
\end{equation}
The eigenvalues $ E_0 < E_1 < E_2$ as functions of $\alpha$ can be viewed in Fig. 2b.

Next we consider subspace(s)
\begin{equation}
\frac{1}{\sqrt{2}} B_\tau^\dagger |0 \rangle , \frac{1}{2 \sqrt{2}} (B_\tau^\dagger)^2  B_{-\tau}^\dagger |0 \rangle ,
\end{equation}
where $\tau = \pm$. In each subspace we have the following eigenvalue problem:
\begin{equation}
\left[
  \begin{array}{cc}
   2 (E_{00} - V_1 + V_2) & - 4 V_2 \\
   - 4 V_2 & 6(E_{00} - V_1) + 8 V_2  \\
  \end{array}
\right],
\end{equation}
and the resulting eigenvalues $E_{7,8} = 2 [2(E_{00} - V_1) + 3 V_2 \mp \sqrt{(E_{00} - V_1 + V_2)^2 + 4 V_2^2}]$.

If we introduce $ X_\tau =   c_{\tau \uparrow}^\dagger d_{\tau \uparrow}^\dagger - c_{\tau \downarrow}^\dagger d_{\tau \downarrow}^\dagger$ we can consider states,
\begin{equation}
\frac{1}{\sqrt{2}} X_\tau^\dagger |0 \rangle ,
\end{equation}
which are eigenvectors with eigenvalue $ E_5 = 2 (E_{00} - V_1)$, and
\begin{equation}
\frac{1}{2 \sqrt{2}} B_\tau^\dagger  B_{-\tau}^\dagger X_\tau^\dagger |0 \rangle ,
\end{equation}
with eigenvalue $ E_6 = 2 [3(E_{00} - V_1) + 2 V_2 ]$.

The subspace(s) described by Eq.(\ref{SS1}) in this case consists of the vectors,
\begin{equation}
c_{\tau \sigma}^\dagger d_{-\tau \sigma'}^\dagger |0 \rangle , \frac{1}{2} B_\tau^\dagger  B_{-\tau}^\dagger c_{\tau \sigma}^\dagger d_{-\tau \sigma'}^\dagger |0 \rangle ,
\end{equation}
where $\sigma , \sigma' = \uparrow , \downarrow$. In each subspace the eigenproblem  is given by the following matrix,
\begin{equation}
\left[
  \begin{array}{cc}
   2 (E_{00} - V_1) & - 2 V_2 \\
   - 2 V_2 & 6(E_{00} - V_1) + 4 V_2  \\
  \end{array}
\right].
\end{equation}
The eigenvalues are $E_{3,4} = 2 [2(E_{00} - V_1) +  V_2 \mp \sqrt{(E_{00} - V_1 + V_2)^2 +  V_2^2}]$, and are plotted in Fig. 2b. The intravalley exciton state(s), $c_{\tau \sigma}^\dagger d_{\tau \sigma'}^\dagger |0 \rangle,$, with
$\sigma \neq \sigma'$, is an eigenstate with eigenvalue $ E_5 = 2 (E_{00} - V_1)$.

Also possible eigenvectors are $X_\tau^\dagger  X_{-\tau}^\dagger  |0 \rangle $ with eigenvalue $ E_9 = 4 (E_{00} - V_1)$, then $B_\tau^\dagger  X_{-\tau}^\dagger  |0 \rangle $, and $(B_\tau^\dagger)^2  |0 \rangle $ (equivalent to
$(X_\tau^\dagger)^2  |0 \rangle $) with common eigenvalue $ E_{10} = 4 (E_{00} - V_1 + V_2)$.

With a single intravalley exciton with different flavors (spins) and $X$'s and $B$'s, we have states
$ B_{-\tau}^\dagger c_{\tau \sigma}^\dagger d_{\tau \sigma'}^\dagger |0 \rangle $ where $\sigma \neq \sigma'$ with
eigenvalue $ E_{10} = 4 (E_{00} - V_1 + V_2)$, and $ X_{-\tau}^\dagger c_{\tau \sigma}^\dagger d_{\tau \sigma'}^\dagger |0 \rangle $ with eigenvalue $ E_9 = 4 (E_{00} - V_1)$, $ (B_{-\tau}^\dagger)^2 c_{\tau \sigma}^\dagger d_{\tau \sigma'}^\dagger |0 \rangle $ (equivalent to $ (X_{-\tau}^\dagger)^2 c_{\tau \sigma}^\dagger d_{\tau \sigma'}^\dagger |0 \rangle ) $ with eigenvalue $ E_6 = 2 [3(E_{00} - V_1) + 2 V_2 ]$, and with a single intervalley exciton we have states $ B_{\tau}^\dagger c_{\tau \sigma}^\dagger d_{-\tau \sigma'}^\dagger |0 \rangle $ and $ B_{-\tau}^\dagger c_{\tau \sigma}^\dagger d_{-\tau \sigma'}^\dagger |0 \rangle $ with energy $ E_{11} = 4 (E_{00} - V_1) + 2 V_2$.
(These states are equivalent to $ X_{\tau}^\dagger c_{\tau \sigma}^\dagger d_{-\tau \sigma'}^\dagger |0 \rangle $ and $ X_{-\tau}^\dagger c_{\tau \sigma}^\dagger d_{-\tau \sigma'}^\dagger |0 \rangle $.)

Now we have to classify remaining 4-particle states that cannot be described solely by $B_\tau, X_\tau$ with $\tau = \pm$, or with a presence of a single intervalley or a special (combining different flavors) intravalley exciton.
They are: $ c_{\tau \sigma}^\dagger d_{\tau \sigma'}^\dagger c_{-\tau \sigma}^\dagger d_{-\tau \sigma'}^\dagger |0 \rangle $, $ c_{\tau \sigma}^\dagger d_{\tau \sigma'}^\dagger c_{-\tau \sigma'}^\dagger d_{-\tau \sigma}^\dagger |0 \rangle $, $ c_{\tau \sigma}^\dagger d_{-\tau \sigma}^\dagger c_{\tau \sigma'}^\dagger d_{-\tau \sigma'}^\dagger |0 \rangle $ with $\sigma \neq \sigma'$ and energy $ E_9 = 4 (E_{00} - V_1)$.

\end{document}